\documentclass[12pt]{article}
\usepackage{graphicx}

\usepackage{epsf}
\def\beq{\begin{equation}}
\def\eeq{\end{equation}}
\def\bea{\begin{eqnarray}}
\def\eea{\end{eqnarray}}
\def\ve{\vert}
\def\vel{\left|}
\def\ver{\right|}
\def\nnb{\nonumber}
\def\ga{\left(}
\def\dr{\right)}

\def\rar{\rightarrow}
\def\nnb{\nonumber}
\def\la{\langle}
\def\ra{\rangle}
\def\ba{\begin{array}}
\def\ea{\end{array}}
\def\bea{\begin{eqnarray}}
\def\eea{\end{eqnarray}}

\def\Bgll{$B_s \rar \gamma \, \ell^+ \ell^-$}

\def\ve{\vert}
\def\vel{\left|}
\def\ver{\right|}
\def\nnb{\nonumber}
\def\ga{\left(}
\def\dr{\right)}

\def\rar{\rightarrow}
\def\nnb{\nonumber}
\def\la{\langle}
\def\ra{\rangle}
\def\lla{\left<}
\def\rra{\right>}
\def\simlt{\stackrel{<}{{}_\sim}}
\def\simgt{\stackrel{>}{{}_\sim}}
\begin{document}
\title{ {\Large {\bf
A General analysis of  the lepton polarizations in  $B_s \rar \gamma \, \ell^+ \ell^-$ decays }}}
\author{ {\small U. O. Y{\i}lmaz$^1$ \,,
B. B.  \c{S}irvanl{\i}$^2$  \,\, and \,\,
G. Turan$^1$}\\
{\small $^1$ Physics Department, Middle East Technical University} \\
{\small 06531 Ankara, Turkey}\\
{\small $^2$ Gazi University, Faculty of Arts and Science, Department of Physics} \\
{\small 06100, Teknikokullar Ankara, Turkey}}

\begin{titlepage}
\maketitle
\thispagestyle{empty}
\begin{abstract}
We present a general analysis of the lepton polarizations in the rare
$B_s \rar \gamma \,\ell^+ \ell^- $ decays by using the most general model independent form of
the effective Hamiltonian.  The sensitivity of the longitudinal, transverse and normal polarizations
of final state leptons, as well as lepton-antilepton combined asymmetries, on the new Wilson coefficients are
investigated. It has been shown that all these physical observables  are very sensitive
to the existence of new physics beyond the standard model  and their experimental measurements can give valuable information
about it.

\noindent PACS number(s): 12.60.Fr, 13.20.He
\end{abstract}
\end{titlepage}
\section{Introduction \label{s1}}
It has been already pointed out many times before \cite{Hurth}
that the rare B meson decays, as being  flavor changing neutral
current (FCNC) processes, are sensitive to the structure of the
standard model (SM),  and its  possible extensions. Therefore,
these decays may  serve as an important tool to investigate the
new physics prior to any possible experimental clue about it. The
experimental situation concerning B physics is promising too. In
addition to several experiments running successfully like the
BELLE experiment at KEK  and the BaBar at SLAC,
new facilities will also start to explore B physics in a near future, like the LHC-B
experiment at CERN and BTeV at FERMILAB.

Among the rare B-meson decays, the semileptonic $B_s \rar \gamma
\, \ell^+ \ell^-$   $(\ell =e,\mu ,\tau)$ decays  are especially
interesting due to their relative cleanliness and sensitivity to
new physics. \Bgll decay is induced by $B\rightarrow \, \ell^+
\ell^-$ one, which can be in principle serve as a useful process
to determine the fundamental parameters of the SM since the  only
non-perturbative quantity in its theoretical calculation  is the
decay constant $f_{B_s}$, which is reliably known. However, in the
SM, matrix element of $B\rightarrow \, \ell^+ \ell^-$ decay is
proportional to the lepton mass and therefore  corresponding
branching ratio  will be  helicity suppressed. Although $\ell
=\tau$ channel is free from this suppression, its experimental
observation is quite difficult due to low efficiency. In this
connection, it has been pointed out \cite{Burdman}-\cite{Korchemsky} that the radiative leptonic
$B^+ \rightarrow \ell^+ \, \nu_{\ell} \, \gamma$ $(\ell =e , \mu)$
decays have larger branching ratios than purely leptonic modes. It has been shown
\cite{Aliev2,Eilam2} that similar enhancements  take place also in the
radiative decay \Bgll , in which the photon emitted from any of
the charged lines in addition to the lepton pair makes it possible
to overcome the helicity suppression. For that reason, the
investigation of the \Bgll decays becomes interesting.

As an exclusive process, the theoretical calculation of \Bgll
decay requires the additional knowledge about the decay form
factors. These are the matrix elements of the effective Hamiltonian between
the initial B and final photon states, when a photon is released
from the initial quark lines, which give rise to the so called
"structure dependent" (SD) contributions to the amplitude, and
between the B and the vacuum states for the "internal
Bremsstrahlung" (IB) part, which arises when a photon is radiated
from final leptons. Finding these hadronic transition matrix
elements  is related to the nonperturbative sector of the QCD and
should be calculated by means of a nonperturbative approach. Thus, their theoretical 
calculation yields the main uncertainty in the prediction of the exclusive rare decays.
The form factors for B decays into $\gamma$ and a vacuum state  have
been calculated in the framework of light-cone QCD sum rules in
\cite{ Eilam1,Aliev2} and in the framework of the light front
quark model in \cite{Geng2}. In addition, it has
been  proposed a model  in \cite{Kruger} for the $B\rightarrow \gamma$ form factors
which  obey all the restrictions   obtained from the  gauge
invariance combined with the large energy effective theory.

Various kinematical distributions of the \Bgll decays have been
studied in many earlier works. The analiysis    in the
framework of the SM can be found in \cite{Aliev2,Aliev3,Eilam2,Geng2}. The new physics effects in
these decays have been studied in some models, like minimal
supersymmetric Standard model (MSSM) \cite{Xiong}-\cite{Choud3}
and the two Higgs doublet model \cite{Iltan1}-\cite{Erkol2}, and
shown that  different observables, like branching ratio,
forward-backward asymmetry, etc., are very sensitive to the
physics beyond the SM. In \Bgll decay, in addition to the
branching ratio and lepton pair forward-backward asymmetry, it is
possible to study some other experimentally observable quantities
associated with the final state leptons and photon, such as the
photon and lepton polarization asymmetries. Along this line, the
polarization asymmetries of the final state lepton in $B_s \rar
\gamma \ell^+ \ell^- $ decays have been studied in MSSM in
\cite{Choud3} and concluded that they can be very useful for
accurate determination of various Wilson coefficients. In
addition, in a recent work \cite{OBG1} we have been considered the
effects of polarized photon in the \Bgll decay  and shown that its
spectrum is sensitive to the new physics effects.

In this work, we will investigate the new physics effects in the
lepton polarization asymmetries in the \Bgll decay. Final state
leptons in the \Bgll decay can have longitudinal $P_L$, transverse
$P_T$ and normal $P_N$ polarizations, where $P_T$ is the
component of the polarization lying in the decay plane and
$P_N$ is the one that is normal to the decay plane. Since these
three components contain different combinations of Wilson
coefficients and hence provide independent information they are
thought to play important role in further investigations of the SM
and its possible extensions. As for the new physics effects,  in rare $B$ meson
decays they can appear in two different ways: one way is
through new contributions to the Wilson coefficients that is
already present in the SM, and the other is through the new
operators in the effective Hamiltonian which is absent in the SM.
In this work we use a most general model independent  effective
Hamiltonian that combines both these approaches and contains the
scalar and tensor type interactions as well as the vector types
(see Eq.(\ref{effH}) below).

The paper is organized as follows: In Sec. \ref{s2},  we first
give the effective Hamiltonian for the quark level process $b \rar  s  \ell^+ \ell^- $
and the  definitions of the form factors, and then introduce the corresponding matrix element.
In Secs. \ref{s3} and \ref{s4}, we present the analytical expressions of the various lepton
polarization asymmetries and lepton-antilepton combined asymmetries, respectively.
Sec. \ref{s5} is devoted to the numerical analysis and discussion of our
results.
\section{Effective Hamiltonian  \label{s2}}
For the radiative \Bgll decay, the basic quark level process is $b \rar  s  \ell^+ \ell^- $,
which can be written in terms of twelve model independent four-Fermi
interactions as follows \cite{Aliev5}:
\bea \label{effH}
{\cal H}_{eff} &=& \frac{\alpha \, G}{\sqrt{2} \,  \pi}
 V_{ts}V_{tb}^\ast
\Bigg\{ C_{SL} \, \bar s i \sigma_{\mu\nu} \frac{q^\nu}{q^2}\, L \,b
\, \bar \ell \gamma^\mu \ell + C_{BR}\, \bar s i \sigma_{\mu\nu}
\frac{q^\nu}{q^2} \,R\, b \, \bar \ell \gamma^\mu \ell \nnb \\
&&+C_{LL}^{tot}\, \bar s_L \gamma_\mu b_L \,\bar \ell_L \gamma^\mu \ell_L +
C_{LR}^{tot} \,\bar s_L \gamma_\mu b_L \, \bar \ell_R \gamma^\mu \ell_R +
C_{RL} \,\bar s_R \gamma_\mu b_R \,\bar \ell_L \gamma^\mu \ell_L \nnb \\
&&+C_{RR} \,\bar s_R \gamma_\mu b_R \, \bar \ell_R \gamma^\mu \ell_R +
C_{LRLR} \, \bar s_L b_R \,\bar \ell_L \ell_R +
C_{RLLR} \,\bar s_R b_L \,\bar \ell_L \ell_R \\
&&+C_{LRRL} \,\bar s_L b_R \,\bar \ell_R \ell_L +
C_{RLRL} \,\bar s_R b_L \,\bar \ell_R \ell_L+
C_T\, \bar s \sigma_{\mu\nu} b \,\bar \ell \sigma^{\mu\nu}\ell \nnb \\
&&+i C_{TE}\,\epsilon^{\mu\nu\alpha\beta} \bar s \sigma_{\mu\nu} b \,
\bar \ell \sigma_{\alpha\beta} \ell  \Bigg\}~, \nnb
\eea
where  $L=(1-\gamma_5 )/2$ and $R=(1+\gamma_5 )/2$  are the chiral projection operators.
In Eq. (\ref{effH}),
$C_X$ are the coefficients of the four--Fermi interactions with $X=LL,LR,RL,RR$
describing vector, $X=LRLR,RLLR,LRRL,RLRL$ scalar and $X=T,TE$ tensor type interactions.
We note that the coefficients $C_{SL}$ and $C_{BR}$ correspond to $-2 m_s C_7^{eff}$ and $-2
m_b C_7^{eff}$ in the SM,  while $C_{LL}$ and $C_{LR}$ are in the form
$C_9^{eff} - C_{10}$ and $C_9^{eff} + C_{10}$, respectively. Therefore, writing
\bea
C_{LL}^{tot} &=& C_9^{eff}-C_{10} + C_{LL}~, \nnb \\
C_{LR}^{tot} &=& C_9^{eff}+C_{10} + C_{LR}~, \nnb
\eea
we observe that $C_{LL}^{tot}$ and $C_{LR}^{tot}$ contain the contributions from the SM and also
from the new physics.

Having established the general form of the effective Hamiltonian,
the next step is to calculate
the matrix element of the $B_s \rar \gamma \, \ell^+\ell^-$ decay, which can be
written as a sum of the SD and the IB parts:
\beq
{\cal M}={\cal M}_{SD}+{\cal M}_{IB}.
\eeq
Using the expressions \cite{Eilam1,Aliev2}
\bea
 \label{mel1}
 \lla \gamma(k) \vel
\bar s \gamma_\mu (1 \mp \gamma_5) b \ver B(p_B) \rra &=&
\frac{e}{m_B^2} \Big\{ \epsilon_{\mu\nu\lambda\sigma}
\varepsilon^{\ast\nu} q^\lambda
k^\sigma g(q^2) \nnb \\
&&\pm i\,
\Big[ \varepsilon^{\ast\mu} (k q) -
(\varepsilon^\ast q) k^\mu \Big] f(q^2) \Big\}~,\\ \nnb \\
\label{mel2}
\lla \gamma(k) \vel \bar s \sigma_{\mu\nu} b \ver B(p_B) \rra &=&
\frac{e}{m_B^2}
\epsilon_{\mu\nu\lambda\sigma} \Big[
G \varepsilon^{\ast\lambda} k^\sigma +
H \varepsilon^{\ast\lambda} q^\sigma +
N (\varepsilon^\ast q) q^\lambda k^\sigma \Big]~,
\eea
\bea
\label{mel1q}
\lla \gamma(k) \vel \bar s (1 \mp \gamma_5) b \ver B(p_B) \rra
&=& 0~, \\
\label{mel2q}
\lla \gamma (k)\vel \bar s i \sigma_{\mu\nu} q^\nu b \ver B (p_B) \rra &=&
\frac{e}{m_B^2} i\, \epsilon_{\mu\nu\alpha\beta} q^\nu
\varepsilon^{\alpha\ast} k^\beta G~,
\eea
and
\bea
\label{mel22q}
\lla \gamma (k) \vel \bar s i \sigma_{\mu\nu} q^\nu (1+\gamma_5) b \ver B (p_B)\rra &=&
\frac{e}{m_B^2} \Big\{
\epsilon_{\mu\alpha\beta\sigma} \, \varepsilon^{\alpha\ast} q^\beta k^\sigma
g_1(q^2)
+ i\,\Big[\varepsilon_\mu^\ast (q k) - (\varepsilon^\ast q) k_\mu \Big]
f_1(q^2) \Big\}~, \nnb \\ & &
\eea
the SD part of the amplitude can be written as
\bea
\label{sd}
{\cal M}_{SD} &=& \frac{\alpha G_F}{4 \sqrt{2} \, \pi} V_{tb} V_{ts}^*
\frac{e}{m_B^2} \,\Bigg\{
\bar \ell \gamma^\mu (1-\gamma_5) \ell \, \Big[
A_1 \epsilon_{\mu \nu \alpha \beta}
\varepsilon^{\ast\nu} q^\alpha k^\beta +
i \, A_2 \Big( \varepsilon_\mu^\ast (k q) -
(\varepsilon^\ast q ) k_\mu \Big) \Big] \nnb \\
&+& \bar \ell \gamma^\mu (1+\gamma_5) \ell \, \Big[
B_1 \epsilon_{\mu \nu \alpha \beta}
\varepsilon^{\ast\nu} q^\alpha k^\beta
+ i \, B_2 \Big( \varepsilon_\mu^\ast (k q) -
(\varepsilon^\ast q ) k_\mu \Big) \Big] \nnb \\
&+& i \, \epsilon_{\mu \nu \alpha \beta}
\bar \ell \sigma^{\mu\nu}\ell \, \Big[ G \varepsilon^{\ast\alpha} k^\beta
+ H \varepsilon^{\ast\alpha} q^\beta +
N (\varepsilon^\ast q) q^\alpha k^\beta \Big] \\
&+& i \,\bar \ell \sigma_{\mu\nu}\ell \, \Big[
G_1 (\varepsilon^{\ast\mu} k^\nu - \varepsilon^{\ast\nu} k^\mu) +
H_1 (\varepsilon^{\ast\mu} q^\nu - \varepsilon^{\ast\nu} q^\mu) +
N_1 (\varepsilon^\ast q) (q^\mu k^\nu - q^\nu k^\mu) \Big] \Bigg\}~,\nnb
\eea
where
\bea
A_1 &=& \frac{1}{q^2} \Big( C_{BR} + C_{SL} \Big) g_1 +
\Big( C_{LL}^{tot} + C_{RL} \Big) g ~, \nnb \\
A_2 &=& \frac{1}{q^2} \Big( C_{BR} - C_{SL} \Big) f_1 +
\Big( C_{LL}^{tot} - C_{RL} \Big) f ~, \nnb \\
B_1 &=& \frac{1}{q^2} \Big( C_{BR} + C_{SL} \Big) g_1 +
\Big( C_{LR}^{tot} + C_{RR} \Big) g ~, \nnb \\
B_2 &=& \frac{1}{q^2} \Big( C_{BR} - C_{SL} \Big) f_1 +
\Big( C_{LR}^{tot} - C_{RR} \Big) f ~, \nnb \\
G &=& 4 C_T g_1 ~~~,~~~ N = - 4 C_T \frac{1}{q^2} (f_1+g_1) ~, \nnb \\
H &=& N (qk) ~~~,~~~G_1 = - 8 C_{TE} g_1 ~, \nnb \\
N_1 &=& 8 C_{TE} \frac{1}{q^2} (f_1+g_1) ~~~,~~~ H_1 = N_1(qk)~ .\nnb
\eea
Here, $\varepsilon_\mu^\ast$ and $k_\mu$ are the four vector polarization
and four momentum of the photon, respectively, $q$ is the momentum transfer,
 $p_B$ is the momentum of the $B$ meson, and $G,~H$ and $N$  have been
expressed  in terms of the form factors $g_1$ and $f_1$ by using Eqs. (\ref{mel2}),
(\ref{mel2q}) and (\ref{mel22q}).

When photon is radiated from the lepton line we get the the so-called "internal  Bremsstrahlung" (IB) contribution,
${\cal M}_{IB}$. Using the expressions
\bea
\la 0 \ve \bar s \gamma_\mu \gamma_5 b \ve B (p_B)\ra &=&
-~i f_B p_{B\mu}~, \nnb \\
\la 0 \ve \bar s \sigma_{\mu\nu} (1+\gamma_5) b \ve B (p_B) \ra &=& 0~,\nnb
\eea
and conservation of the vector current, we get
\bea
\label{ib}
{\cal M}_{IB} &=& \frac{\alpha G_F}{4 \sqrt{2} \, \pi} V_{tb} V_{ts}^*
e f_B i \,\Bigg\{
F\, \bar \ell  \Bigg(
\frac{{\not\!\varepsilon}^\ast {\not\!p}_B}{2 p_1 k} -
\frac{{\not\!p}_B {\not\!\varepsilon}^\ast}{2 p_2 k} \Bigg)
\gamma_5 \ell \nnb \\
&+& F_1 \, \bar \ell  \Bigg[
\frac{{\not\!\varepsilon}^\ast {\not\!p}_B}{2 p_1 k} -
\frac{{\not\!p}_B {\not\!\varepsilon}^\ast}{2 p_2 k} +
2 m_\ell \Bigg(\frac{1}{2 p_1 k} + \frac{1}{2 p_2 k}\Bigg)
{\not\!\varepsilon}^\ast \Bigg] \ell \Bigg\}~,
\eea
where $p_1$ and $p_2$ are the momenta of the $\ell^-$ and $\ell^+$, respectively, and
\bea
F &=& 2 m_\ell \Big( C_{LR}^{tot} - C_{LL}^{tot} + C_{RL} - C_{RR} \Big)
+ \frac{m_B^2}{m_b}
\Big( C_{LRLR} - C_{RLLR} - C_{LRRL} + C_{RLRL} \Big)~, \nnb \\
F_1 &=&\frac{m_B^2}{m_b} \Big( C_{LRLR} - C_{RLLR} + C_{LRRL} - C_{RLRL}
\Big)~.
\eea

The next task is the calculation of the  differential decay rate
of $B_s \rar \gamma \, \ell^+ \ell^- $ decay  as a function of
dimensionless parameter $1-s=1-2 E_{\gamma}/m_B$, where $E_{\gamma}$
is the  photon energy. In the center of mass (CM) frame of the
dileptons $\ell^+\ell^-$, where we take  $z=\cos \theta $ and $\theta$ is the angle
between the momentum of the $B_{s}$-meson and that of $\ell^-$, double differential decay width
is found to be
\bea
\label{dGdxdz}
\frac{d \Gamma}{ds \, dz} = \frac{1}{(2 \pi)^3 64 }\, (s-1) \, v \, m_B \, \vel {\cal M} \ver^2~,
\eea
with
\bea
\vel {\cal M} \ver^2 & = &\vel {\cal M}_{SD} \ver^2+\vel {\cal M}_{IB} \ver^2+
2 Re({\cal M}_{SD}{\cal M}^{*}_{IB} )\label{M2}
\eea
where $v=\sqrt{1-\frac{4 r}{s}}$ and $r=m^2_{\ell}/m^2_{B}$.

\section{Lepton polarization asymmetries \label{s3}}
Now, we would like to discuss the  lepton polarizations in the rare \Bgll decays.
For $i=L,~T,~N$, the polarization asymmetries $P^{\mp}_{i}$ of the final $\ell^{\mp}$ lepton
are defined as
\begin{eqnarray}
P^{\mp}_{i} (s) & = & \frac{\frac{d\Gamma}{ds} (\vec{n}^{\mp}=\vec{e}^{\,\mp}_i)-
\frac{d\Gamma}{ds} (\vec{n}^{\mp}=-\vec{e}^{\,\mp}_i)}
{\frac{d\Gamma}{ds} (\vec{n}^{\mp}=\vec{e}^{\,\mp}_i)+
\frac{d\Gamma}{ds} (\vec{n}^{\mp}=-\vec{e}^{\,\mp}_i)} \label{PL}\,,
\end{eqnarray}
where $\vec{n}^{\mp}$ is the unit vectors in the $\ell^{\mp}$ rest frame, which are defined as
\bea
S^{-\mu}_L&\equiv&(0,\vec{e}^{\,-}_L)=\Bigg(0,\frac{\vec{p}_1}{|\vec{p}_1|}\Bigg)\, ,\nnb\\
S^{-\mu}_N&\equiv&(0,\vec{e}^{\,-}_N)=\Bigg(0,\frac{\vec{k}\times\vec{p}_1}{|\vec{k}\times\vec{p}_1|}\Bigg) \, ,\nnb\\
S^{-\mu}_T&\equiv&(0,\vec{e}^{\,-}_T)=\Bigg(0,\vec{e}^{\,-}_N\times \vec{e}^{\,-}_L \Bigg)\,, \nnb \\
S^{+\mu}_L&\equiv&(0,\vec{e}^{\,+}_L)=\Bigg(0,\frac{\vec{p}_2}{|\vec{p}_2|}\Bigg) \, ,\nnb\\
S^{+\mu}_N&\equiv&(0,\vec{e}^{\,+}_N)=\Bigg(0,\frac{\vec{k}\times\vec{p}_2}{|\vec{k}\times\vec{p}_2|}\Bigg) \, , \nnb\\
S^{+\mu}_T&\equiv&(0,\vec{e}^{\,+}_T)=\Bigg(0,\vec{e}^{\,+}_N\times\vec{e}^{\,+}_L \Bigg)\,.
\eea
The longitudinal unit vector
$S_L$ is boosted to the CM frame of $\ell^{+}\ell^{-}$ by Lorentz
transformation:
\bea
S^{-\mu}_{L,CM}& = & \Bigg(\frac{|\vec{p}_1|}{m_\ell},\frac{E_\ell~\vec{p}_1}{m_\ell|\vec{p}_1|}\Bigg) \, ,\nnb \\
S^{+\mu}_{L,CM}& = & \Bigg(\frac{|\vec{p}_1|}{m_\ell},-\frac{E_\ell~\vec{p}_1}{m_\ell|\vec{p}_1|}\Bigg) \, ,
\eea
while $P_T$ and $P_N$ are not changed by the boost since they lie in the  perpendicular directions.

After some lengthy algebra, we obtain the following expressions
for the polarization components of the $\ell^\pm$ leptons in \Bgll decays:
\bea \label{PLmp}
P^{\pm}_{L} &=&\frac{1}{6 v\, \Delta_0} \Bigg\{(1-s) v^{3}
\Bigg(\pm\frac{m_{B}^{3} (s-1)^{2} s (12 r + s(v^{2}-1))
\mbox{\rm Im}[(A_2-B_2) N_1^\ast]}{\sqrt{r}}\nnb \\
&+& 4 m_{B}^{2} (s-1)^{2} s \Big(\pm (\vel A_1 \ver^2+ \vel A_2
\ver^2- \vel B_1 \ver^2- \vel B_2 \ver^2 )-\mbox{\rm Im}[G
N_1^\ast]+\mbox{\rm Im}[G_1 N^\ast]\Big)\nnb \\
&+& 24 (s-1) s (\mbox{\rm Im}[G_1 H^\ast]-\mbox{\rm Im}[G H_1^\ast] )
+ 4s^{2} (-12 \mbox{\rm Im}[H_1 H^\ast]\nnb \\
&+& m_{B}^{4} (s-1)^{2} \mbox{\rm Im}[N_1 N^\ast]) +
16(s-1)^{2} \mbox{\rm Im}[(-G\mp m_{B}\sqrt{r}A_2) G_1^\ast]\nnb \\
& \pm & \frac{m_{B}(s-1)^{2} (-12r+s(v^{2}-1))(\mbox{\rm Im}[B_2 G_1^\ast] -
\mbox{\rm Re}[(-A_1+B_1) G^\ast])}{\sqrt{r}}\nnb \\
&-& \frac{m_{B}(s-1)^{2} (12r+s(v^{2}-1))(\mbox{\rm Im}[(A_1+B_1)
G_1^\ast] + \mbox{\rm Re}[(A_2+B_2) G^\ast])}{\sqrt{r}}\nnb \\
&+& 24 m_{B}\sqrt{r}s (s-1) (\mp2\mbox{\rm Im}[( B_2-A_2) H_1^\ast]
+ \mbox{\rm Re}[(A_2+B_2) H^\ast])\nnb \\
&-& \frac{m_{B}^{3}(s-1)^{2} s^{2}(v^{2}-1)\mbox{\rm Re}[(A_2+B_2)
N^\ast]}{\sqrt{r}}\Bigg)\nnb \\
&-& \frac{48f_{B}^{2}(1+s^{2}-4r(1+s))(s v + (2r-s)\mbox{\rm
ln}[u])\mbox{\rm Re}[F_1 F^\ast]}{(s-1)s}\nnb \\
&+& 24 f_{B} \mbox{\rm ln}[u] \Bigg( 2 (-s+2r (1+s))\mbox{\rm Im}[F H_1^\ast]
\mp \frac{ m_{l}}{s} (2r-s) (s-1)^{2}(\mbox{\rm
Re}[(A_1-B_1) F^\ast]\nnb \\
&- & \mbox{\rm Re}[(A_2- B_2) F_1^\ast]) - \frac{2m_{l}r}{s}  (s-1) ((s-1)\mbox{\rm
Re}[(A_1+B_1) F_1^\ast]\nnb \\
&+& (1+s)\mbox{\rm Re}[(A_2+B_2) F^\ast]) - \frac{4r}{s}  (s-1)(\mbox{\rm Im}[F G_1^\ast] +
(4r-1)\mbox{\rm Re}[F_1 G^\ast])\nnb \\
&-& 2(4r-1) s v^{2} \mbox{\rm Re}[F_1 H^\ast] + m_{B}^{2} (s-1) (-s \mbox{\rm Im}[F
N_1^\ast] + (s-2r(1+s)) \mbox{\rm Re}[F_1 N^\ast])\Bigg)\nnb \\
&+& 24 f_{B} v \Bigg(2 (s-1) \mbox{\rm Im}[F (G_1 + m_{B}^{2}N_1 - s
H_1)^\ast]- m_{l}(s-1)^{2}\mbox{\rm Re}[(A_1\mp A_2+B_1) F_1^\ast] \nnb \\
& \mp &  m_{l} (s-1)^{2}\mbox{\rm Re}[A_1 F^\ast + B_2 F_1^\ast]
+ (s-1)\Big(-2(1-4r)\mbox{\rm Re}[F_1 G^\ast] \nnb \\
& +& m_{l}\mbox{\rm Re}[\Big(-(s+1) A_2 \pm (s-1)B_1 \Big)F^\ast ]\Big)
+ (1-s) ( m_{l}(1+s)\mbox{\rm Re}[B_2 F^\ast] \nnb \\
&-& 2 s v^{2} \mbox{\rm Re}[F_1 H^\ast] +
2m_{B}^{2} (s-2r(1+s))\mbox{\rm Re}[F_1 N^\ast])\Bigg)\Bigg\}\, ,
\eea
\bea \label{PTmp}
P^{\pm}_{T} &=&\frac{1}{\Delta_0}\Bigg\{
 \frac{(2\sqrt{r}-\sqrt{s})}{s v}  (1-s)  f_{B} m_{B}\pi \Bigg( \pm s v^2 (1+s)
\mbox{\rm Re}[(A_1-B_1) F^\ast]\nnb \\
&+& (s-1)(4 r+s)\mbox{\rm Re}[(A_2+B_2)F^\ast]+
(4r (1-3 s)+s (s+1))\mbox{\rm Re}[(A_1+B_1)F_1^\ast]\nnb \\
& \pm & s v^2 (s-1)\mbox{\rm Re}[(A_2-B_2)F_1^\ast] -  8\sqrt{r}
(\mbox{\rm Im}[F (G_1(s-1)-2 s H_1)^\ast]-(1-4r)\mbox{\rm Re}[F_1 G^\ast])/m_B \Bigg)\nnb \\
&+& \frac{\pi v}{4\sqrt{s}}(s-1)^{2} \Bigg(8\sqrt{r}\, \mbox{\rm Im}[(G_1(s-1)+2sH_1) G^\ast] +
 2 m_{B} s (-(4r+s) \mbox{\rm Im}[(A_1+B_1) H_1^\ast]\nnb \\
&\mp & (4r-s) \mbox{\rm Re}[( A_1 - B_1) H^\ast]) - 2 m_{B}^{2} \sqrt{r} (s-1) s \,
\mbox{\rm Re}[(A_1+B_1) (A_2+B_2)^\ast]\nnb \\
&+ & m_{B} (s-1) \Big(\mp(s-4 r) (\mbox{\rm Im}[(A_2-B_2) G_1^\ast]+\mbox{\rm Re}[(A_1-B_1) G^\ast])
+ (4r+s) (\mbox{\rm Im}[(A_1+B_1) G_1^\ast]\nnb \\
&+& \mbox{\rm Re}[(A_2+B_2) G^\ast])\Big)\Bigg) + 4\pi v f_{B}^{2} (4r-1) \mbox{\rm Re}[F_1 F^\ast]\Bigg\} \, ,
\eea
\bea \label{PNmp}
P^{\pm}_{N} &=&  \frac{\pi }{4\Delta_0} (1-s) \Bigg\{
(1-s)\sqrt{s} v^{2} \Big(\pm2 m_{B}^{2} \sqrt{r} (s-1) (\mbox{\rm
Im}[A_1 B_2^\ast]+\mbox{\rm Im}[A_2 B_1^\ast]) + 8\sqrt{r}
(\mbox{\rm Im}[G H^\ast]\nnb \\ &-& \mbox{\rm Im}[G_1 H_1^\ast]) -
2 m_{B} s (\mbox{\rm Im}[(A_1+B_1) H^\ast]\pm \mbox{\rm
Re}[(A_1-B_1) H_1^\ast])\nnb \\ &+& m_{B} (s-1) (\mbox{\rm
Im}[( A_1\mp A_2+ B_1\pm B_2) G^\ast]-\mbox{\rm Re}[(\mp A_1+
A_2\pm B_1+ B_2) G_1^\ast])\Big)\nnb \\ &-&
4  (2\sqrt{r}-\sqrt{s}) m_{B}f_{B} \Big((1+s) \mbox{\rm Im}[(A_1+B_1) F^\ast]
\pm (1+s-8 r)  \mbox{\rm Im}[(A_1-B_1) F_1^\ast]\nnb \\
&\mp &  (1-s)  \mbox{\rm Im}[(A_2-B_2) F^\ast] + (s-1)
\mbox{\rm Im}[(A_2+B_2) F_1^\ast]\nnb \\
&+& 8\sqrt{r} (\mbox{\rm Im}[F (G-H)^\ast]+ \mbox{\rm Re}[F_1
H_1^\ast] )/m_B\Big)\Bigg\}\, ,
\eea
where $u=1+v/1-v$ and
\bea
\label{bela2} \lefteqn{ \Delta_0 = \Bigg\{ (1-s)^3 v\, \Bigg (4
m_\ell \, \mbox{\rm Re}\Big( [A_1+B_1] G^\ast\Big)
- \, 4 m_B^2 r \, \mbox{\rm Re}[ A_1 B_1^\ast + A_2 B_2^\ast ]} \nnb \\
&&-  4 \Big[ \vel H_1 \ver^2 s + \mbox{\rm Re}[ G_1 H_1^\ast ] (1-s) \Big]
\frac{(8 r +s)}{(1-s)^2} -  4  \Big[ \vel H
\ver^2 s + \mbox{\rm Re}[ G H^\ast ] (1-s)
\Big] \frac{(4 r+s)}{(1-s)^2} \nnb \\
&&+ \, \frac{1}{3} m_B^2 \Big[ 2\,\mbox{\rm Re}[ G N^\ast ] +
m_B^2 \vel N \ver^2 s \Big] (s-4r) \nnb \\
&&+ \, \frac{1}{3} m_B^2 \Big[ 2\,\mbox{\rm Re}[G_1 N_1^\ast ] +
m_B^2 \vel N_1 \ver^2 s \Big] (s+8r) \nnb \\
&&- \, \frac{2}{3} m_B^2 \Big( \vel A_1 \ver^2 + \vel A_2 \ver^2 +
\vel B_1 \ver^2 + \vel B_2 \ver^2\Big) ( s - r ) - \frac{4}{3}
\Big( \vel G \ver^2 + \vel G_1 \ver^2 \Big)
\frac{(s + 2 r )}{s}\nnb \\
&&+\,2 m_\ell \,\mbox{\rm Im}\Big([A_2+B_2] [6 H_1^\ast s
 + 2 G_1^\ast (1-s) - m_B^2 \,N_1^\ast  (1-s) s]\Big)\frac{1}{(1-s)} \Bigg ) \nnb \\
&&+ \, 4 f_B \,\Bigg ( 2 v \,\Bigg[ \mbox{\rm Re}[F
G^\ast]\frac{1}{s} - \mbox{\rm Re}[F H^\ast] +
\,m_B^2\, \mbox{\rm Re}[F N^\ast ] + m_\ell \,\mbox{\rm
Re}\Big([A_2+B_2] F_1^\ast\Big)
\Bigg] \, (1-s) s\nnb \\
&&+\, {\rm ln} [u] \Bigg[ m_\ell \, \mbox{\rm Re}[(A_2+B_2)
F_1^\ast] \,(1-s) (1-s-4 r) + 2  \, \mbox{\rm Re}[F H^\ast ]
\Big[s - 2 r (s+1) \Big] \nnb \\
&&- \, 4 r (1-s)\, \mbox{\rm Re}[F G^\ast]
- m_B^2 \, \mbox{\rm Re}[F N^\ast ] \, (1-s) s
- m_\ell \,\mbox{\rm Re}[(A_1+B_1) F^\ast] \, (1-s)^2
\Bigg]\Bigg ) \nnb \\
&&+ \,  4 f_B^2
\Bigg (2 v\, \Big( \vel F \ver^2+
(1-4 r) \vel F_1 \ver^2 \Big) \frac{s}{(1-s)} + \, {\rm ln}[u] \Bigg[ \vel F \ver^2
\Big( 2 +\frac{2(2 r-1)}{(1-s)} -(1-s) \Big)
\nnb \\
&&+ \vel F_1 \ver^2 \Bigg( 2 (1-4 r) - \frac{2 \ga 1- 6 r + 8 r^2 \dr}{(1-s)} -(1-s)
\Bigg) \Bigg]\Bigg )\Bigg\}~.
\eea
From Eqs. (\ref{PLmp})-(\ref{PNmp}), we see that in the limit $m_{\ell}\rightarrow 0$,
longitudinal polarization asymmetry  for the \Bgll decay is only
determined by the scalar and tensor interactions, while transverse and normal components receive
contributions mainly from the tensor and scalar interactions, respectively. Therefore, experimental
measurement of these observables may provide important hints for the new physics beyond the SM.
\section{Lepton-antilepton combined asymmetries \label{s4}}
One can also obtain useful information about new physics by performing a combined analysis of
the lepton and antilepton polarizations.
In an earlier work along this line, the combinations $P^{-}_{L}+P^{+}_{L}$,
$P^{-}_{T}-P^{+}_{T} $ and $P^{-}_{N}+P^{+}_{N}$
were considered for the inclusive
$B \rightarrow X_s \tau^+ \tau^-$  decay \cite{Fukae}, because  it was argued that within the SM $P^{-}_{L}+P^{+}_{L}=0$,
$P^{-}_{T}-P^{+}_{T}\approx 0 $ and $P^{-}_{N}+P^{+}_{N}=0$ so that any deviation from these results
would be a definite indication of new physics. Later  same discussion was done in
connection with the exclusive processes $B \rightarrow K^*,K \ell^+ \ell^-$  and shown that
within the SM the above-mentioned combinations of the $\ell^+$ and $\ell^-$ polarizations  vanish only
at  zero lepton mass limit \cite{Aliev6}. In \cite{Choud1},  the same combinations of the lepton and antilepton
polarizations were analyzed in for \Bgll decay within the MSSM model and concluded that the results quoted
in earlier works that  these quantities identically vanish in the SM was a process dependent statement.

Now, we would like to analyze the same combinations of the various polarization asymmetries in a model
independent way and discuss the possible new physics effects through these observables.

For $P^{-}_{L}+P^{+}_{L}$, we find from Eq. (\ref{PLmp}) that
\bea \label{PLmPLp}
P^{-}_{L}+P^{+}_{L} &=&\frac{1}{3 v\, \Delta_0} \Bigg\{(1-s) v^{3}
\Bigg( 4 m_{B}^{2} (s-1)^{2} s \Big(\mbox{\rm Im}[G_1 N^\ast]-\mbox{\rm Im}[G N_1^\ast]\Big)\nnb \\
&+& 24 (s-1) s (\mbox{\rm Im}[G_1 H^\ast]-\mbox{\rm Im}[G H_1^\ast] )
+ 4s^{2} (-12 \mbox{\rm Im}[H_1 H^\ast]\nnb \\
&+& m_{B}^{4} (s-1)^{2} \mbox{\rm Im}[N_1 N^\ast]) +
16(s-1)^{2} \mbox{\rm Im}[-G G_1^\ast]\nnb \\
&-& \frac{m_{B}(s-1)^{2} (12r+s(v^{2}-1))(\mbox{\rm Im}[(A_1+B_1)
G_1^\ast] + \mbox{\rm Re}[(A_2+B_2) G^\ast])}{\sqrt{r}}\nnb \\
&+& 24 m_{B}\sqrt{r}(s-1)s \mbox{\rm Re}[(A_2+B_2) H^\ast]
-\frac{m_{B}^{3}(s-1)^{2} s^{2}(v^{2}-1)\mbox{\rm Re}[(A_2+B_2)
N^\ast]}{\sqrt{r}}\Bigg)\nnb \\
&-& \frac{48f_{B}^{2}(1+s^{2}-4r(1+s))(s v + (2r-s)\mbox{\rm
ln}[u])\mbox{\rm Re}[F_1 F^\ast]}{(s-1)s}\nnb \\
&+& 24 f_{B} \mbox{\rm ln}[u] \Bigg( 2 (-s+2r (1+s)) \mbox{\rm Im}[F H_1^\ast] -
  \frac{4r}{s}  (s-1)(\mbox{\rm Im}[F G_1^\ast] +
(4r-1)\mbox{\rm Re}[F_1 G^\ast])\nnb \\
& - & \frac{2m_{l}r}{s}  (s-1) \Big((s-1)\mbox{\rm Re}[(A_1+B_1) F_1^\ast]+
(1+s)\mbox{\rm Re}[(A_2+B_2) F^\ast]\Big) \nnb \\
&-& 2(4r-1) s v^{2} \mbox{\rm Re}[F_1 H^\ast] + m_{B}^{2} (s-1) (-s \mbox{\rm Im}[F
N_1^\ast] + (s-2r(1+s)) \mbox{\rm Re}[F_1 N^\ast])\Bigg)\nnb \\
&+& 24 f_{B} v (s-1) \Bigg(2  \mbox{\rm Im}[F (G_1 + s(m_{B}^{2}N_1 -
H_1))^\ast]- m_{l}(s-1)\mbox{\rm Re}[(A_1+B_1) F_1^\ast] \nnb \\
& - & 2\Big((1-4r)\mbox{\rm Re}[F_1 G^\ast]-s v^2 \mbox{\rm Re}[F_1 H^\ast]\Big)
-m_{l}  (1+s)\mbox{\rm Re}[(A_2+B_2) F^\ast ]\nnb \\ & - & 2
(s-2r(1+s))m_{B}^{2} \mbox{\rm Re}[F_1 N^\ast]\Bigg)\Bigg\}\, .
\eea
We now consider $P^{-}_{T}-P^{+}_{T}$. It reads from Eq. (\ref{PTmp}) as
\bea \label{PTmmPTp}
P^{-}_{T}-P^{+}_{T} &=&\frac{2 \pi v}{\Delta_0}m_{B}(s-1)\Bigg\{
(2\sqrt{r}-\sqrt{s}) f_{B}   \Bigg( (s+1)\mbox{\rm Re}[(A_1-B_1)F^\ast]+
(s-1)\mbox{\rm Re}[(A_2-B_2)F_1^\ast] \Bigg)\nnb \\
&+& \frac{ 1}{4\sqrt{s}}(s-1)  \Bigg(
 2  s  (4r-s) \mbox{\rm Re}[( A_1 - B_1) H^\ast]
+ (s-1) \Big((s-4 r) (\mbox{\rm Im}[(A_2-B_2) G_1^\ast]\nnb \\ & + & \mbox{\rm Re}[(A_1-B_1) G^\ast])
\Big)\Bigg) \Bigg\} \, .
\eea
Finally, for $P^{-}_{N}+P^{+}_{N}$, we get from Eq, (\ref{PNmp})
\bea \label{PNmPNp}
P^{-}_{N}+P^{+}_{N} &=&  \frac{\pi   }{2\Delta_0} (1-s)m_{B} \Bigg\{
(1-s)\sqrt{s} v^{2} \Bigg( 8\sqrt{r}(\mbox{\rm Im}[G H^\ast]- \mbox{\rm Im}[G_1 H_1^\ast])/m_B \nnb \\ &-&
2  s (\mbox{\rm Im}[(A_1+B_1) H^\ast])+ (s-1) (\mbox{\rm
Im}[( A_1+ B_1) G^\ast]-\mbox{\rm Re}[(A_2+ B_2) G_1^\ast])\Bigg)\nnb \\ &-&
4 (2\sqrt{r}-\sqrt{s}) f_{B} \Bigg((1+s) \mbox{\rm Im}[(A_1+B_1) F^\ast]
+(s-1) \mbox{\rm Im}[(A_2+B_2) F_1^\ast]\nnb \\
&+& 8\sqrt{r} (\mbox{\rm Im}[F (G-H)^\ast]+ \mbox{\rm Re}[F_1 H_1^\ast] )/m_B\Bigg)\Bigg\}\, .
\eea

We can now  easily obtain from Eq. (\ref{PLmPLp}-\ref{PNmPNp}) that  sum of the longitudinal and
normal polarization asymmetries of $\ell^+$ and $\ell^-$ and the difference of transverse polarization
asymmetry for \Bgll decay do not vanish in the SM, but given by
\bea
(P^{-}_{L}+P^{+}_{L})_{SM} & = & \frac{64 f_B}{s v}m^2_{\ell}(1+s)(1-s)(s v-2 r \mbox{\rm ln}[u])
\mbox{\rm Re}\Big[C_{10} \Big(C^{eff}_9 \, f -\frac{2C^{eff}_7 m_b \, }{q^2}f_1\Big)^{\ast}\Big]
\, ,\nnb \\
(P^{-}_{T}-P^{+}_{T})_{SM} & = & 16 f_B \pi m_{\ell}v (1-s^2) (2 \sqrt{r}-\sqrt{s})
|C_{10}|^2 \, g \, , \nnb \\
(P^{-}_{N}+P^{+}_{N})_{SM} & = & 16 f_B \pi \, m_B m_{\ell} (s+1)(s-1)(2 \sqrt{r}-\sqrt{s})
\mbox{\rm Im}\Big[C_{10} \Big(C^{eff}_9 \, g -\frac{2C^{eff}_7 m_b \, }{q^2}g_1\Big)^{\ast}\Big]
\, ,\nnb \\
\eea
which do not coincide  with those given in \cite{Choud1}, although our conclusion that within the SM,
$P^{-}_{L}+P^{+}_{L}=0$, $P^{-}_{T}-P^{+}_{T}\approx 0 $ and $P^{-}_{N}+P^{+}_{N}=0$ at only
zero lepton mass limit, does.

Before giving our numerical results and their discussion,  we like to note a final point about their
calculations.
As seen from the expressions of the lepton polarizations given by Eqs.(\ref{PLmp}-\ref{PNmPNp}), they
are functions of $s$ as well as the new Wilson coefficients. Thus, in order to investigate the
dependencies of these observables on the new Wilson coefficients, we eliminate the parameter $s$
by performing its integration over the allowed kinematical region. In this way we obtain the
average values of the lepton polarizations, which are defined by
\beq
\langle P_i \rangle = \frac{\int^{1-\delta}_{(2m_{\ell}/m_B)^2}\, P_i(s) \, \frac{d\Gamma}{ds}ds}
{\int^{1-\delta}_{(2m_{\ell}/m_B)^2}\,  \frac{d\Gamma}{ds}ds}\label{Pidef}\, .
\eeq
We note that the part of $d\Gamma/ds$ in (\ref{Pidef}) which receives contribution from the
$\vel {\cal M}_{IB} \ver^2 $ term has infrared singularity due to
the emission of soft photon. To obtain a finite result from these integrations, we follow
the approach described in \cite{Aliev2} and impose a cut on the photon energy,
i.e., we require $E_{\gamma}\geq 25$ MeV, which corresponds to detect only hard photons experimentally.
This cut implies that $E_{\gamma}\geq \delta \, m_B /2$ with $\delta =0.01$.

\section{Numerical analysis and discussion \label{s5}}
We present here our numerical analysis about the averaged polarization asymmetries $<P^-_L>$, $<P^-_T>$
and $<P^-_N>$ of $\ell^-$ for the $B_s \rar \gamma \ell^+ \ell^- $ decays with $\ell =\mu , \tau $, as well as
the lepton-antilepton combined asymmetries $<P^-_L+P^+_L>$, $<P^-_T-P^+_T>$ and $<P^-_N+P^+_N>$.
We first give the input parameters used in our numerical analysis :
\begin{eqnarray}
& & m_B =5.28 \, GeV \, , \, m_b =4.8 \, GeV \, , \,m_{\mu} =0.105 \, GeV \, , \,
m_{\tau} =1.78 \, GeV \, , \nnb \\
& & f_B=0.2 \, GeV \, , \, \, |V_{tb} V^*_{ts}|=0.045 \, \, , \, \, \alpha^{-1}=137  \, \,  ,
G_F=1.17 \times 10^{-5}\, GeV^{-2} \nnb \\
& &  \tau_{B_{s}}=1.54 \times 10^{-12} \, s \, .
\end{eqnarray}
The values of the individual Wilson coefficients that appear in the SM are listed in Table (\ref{table1}).
\begin{table}
        \begin{center}
        \begin{tabular}{|c|c|c|c|c|c|c|c|c|}
        \hline
        \multicolumn{1}{|c|}{ $C_1$}       &
        \multicolumn{1}{|c|}{ $C_2$}       &
        \multicolumn{1}{|c|}{ $C_3$}       &
        \multicolumn{1}{|c|}{ $C_4$}       &
        \multicolumn{1}{|c|}{ $C_5$}       &
        \multicolumn{1}{|c|}{ $C_6$}       &
        \multicolumn{1}{|c|}{ $C_7^{\rm eff}$}       &
        \multicolumn{1}{|c|}{ $C_9$}       &
                \multicolumn{1}{|c|}{$C_{10}$}      \\
        \hline
        $-0.248$ & $+1.107$ & $+0.011$ & $-0.026$ & $+0.007$ & $-0.031$ &
   $-0.313$ &   $+4.344$ &    $-4.624$       \\
        \hline
        \end{tabular}
        \end{center}
\caption{ Values of the SM Wilson coefficients at $\mu \sim m_b $ scale.\label{table1}}
\end{table}

It should be  noted here  that the value of the Wilson coefficient $C_9$ in Table (\ref{table1})
 corresponds
only to the short-distance contributions. $C_9$ also receives long-distance
contributions due to conversion of the real $\bar{c}c$ into
lepton pair $\ell^+ \ell^-$ and they are usually absorbed into a redefinition of the short-distance Wilson
coefficients:
\begin{eqnarray}
C_9^{eff}(\mu)=C_9(\mu)+ Y(\mu)\,\, ,
\label{C9efftot}
\end{eqnarray}
where
\begin{eqnarray}
\label{EqY}
Y(\mu)&=& Y_{reson}+ h(y,s) [ 3 C_1(\mu) + C_2(\mu) +
3 C_3(\mu) + C_4(\mu) + 3 C_5(\mu) + C_6(\mu)] \nonumber \\&-&
\frac{1}{2} h(1, s) \left( 4 C_3(\mu) + 4 C_4(\mu)
+ 3 C_5(\mu) + C_6(\mu) \right)\nnb \\
&- &  \frac{1}{2} h(0,  s) \left[ C_3(\mu) + 3 C_4(\mu) \right]
\\&+& \frac{2}{9} \left( 3 C_3(\mu) + C_4(\mu) + 3 C_5(\mu) +
C_6(\mu) \right) \nonumber \,\, ,
\end{eqnarray}
and $y=m_c/m_b$, $s=q^2/m^2_B\equiv 1-x$ and the functions $h(y,s)$ arises from
the one loop contributions of the four quark operators $O_1$,...,$O_6$ and their explicit
forms can be found in \cite{Misiak}.
It is possible to parametrize  the resonance $\bar{c}c$
contribution $Y_{reson}(s)$ in Eq.(\ref{EqY}) using a Breit-Wigner
shape with normalizations fixed by data which is given by
\cite{AAli2}
\begin{eqnarray}
Y_{reson}(s)&=&-\frac{3}{\alpha^2_{em}}\kappa \sum_{V_i=\psi_i}
\frac{\pi \Gamma(V_i\rightarrow \ell^+
\ell^-)m_{V_i}}{s m^2_B-m_{V_i}+i m_{V_i}
\Gamma_{V_i}} \nonumber \\
&\times & [ (3 C_1(\mu) + C_2(\mu) + 3 C_3(\mu) + C_4(\mu) + 3
C_5(\mu) + C_6(\mu))]\, ,
 \label{Yresx}
\end{eqnarray}
where the phenomenological parameter $\kappa$ is usually taken as
$\sim 2.3$.

As for the values of the new Wilson coefficients, they are the free parameters in this work,
but  it is possible to establish ranges out of experimentally measured branching ratios of the
semileptonic and also purely leptonic rare B-meson decays
\bea
BR (B \rar K \, \ell^+ \ell^-) & = & (0.75^{+0.25}_{-0.21}\pm 0.09) \times 10^{-6} \, \, ,\nnb \\
BR (B \rar K^* \, \mu^+ \mu^-) & = & (0.9 ^{+1.3}_{-0.9}\pm 0.1)\times 10^{-6}\, \, ,\nnb
\eea
reported  by Belle and Babar collaborations \cite{ABE}. It is now also available an upper bound of pure
leptonic rare B-decays in the $B^0 \rar  \mu^+ \mu^-$ mode \cite{Halyo}:
\bea
BR ( B^0 \rar  \mu^+ \mu^-) & \leq & 2.0 \times 10^{-7}  \, \, .\nnb
\eea
Being in accordance with this upper limit and also the above mentioned measurements of the branching ratios
for the semileptonic rare B-decays, we take in this work all new Wilson coefficients as real and varying in
the region $-4\leq C_X\leq 4$.

Among the new Wilson coefficients that appear in Eq.(\ref{effH}), those related to the
helicity-flipped counter-parts of the SM operators, namely, $C_{RL}$ and $C_{RR}$, vanish in all models with
minimal flavor violation in the limit $m_s \rar 0$. However, there are some MSSM scenarios in which
there are finite contributions from these vector operators even for a vanishing s-quark mass. In addition,
scalar type interactions can also contribute through the neutral Higgs diagrams in e.g. multi-Higgs doublet models
and MSSM  for some regions of the parameter spaces of the related models.
In literature there exists studies to establish ranges out of constraints under various precision measurements
for these coefficients (see e.g. \cite{HuangWu}) and our choice for the range of the new Wilson coefficients are
in agreement with these calculations.


To make some numerical predictions, we also need the explicit forms of the form factors $g,~f,~g_1$ and $f_1$.
In our work we have used the results of  \cite{Aliev2}, in which   $q^2$ dependencies of the form factors
are given as
\begin{eqnarray}
g(q^2) & = & \frac{1 \, GeV}{\left(1-\frac{q^2}{5.6^2}\right)^2} \, \, ,\, \,
f(q^2) = \frac{0.8 \, GeV}{\left(1-\frac{q^2}{6.5^2}\right)^2} \, \, , \, \,
g_1(q^2) = \frac{3. 74 \, GeV^2}{\left(1-\frac{q^2}{40.5}\right)^2} \, \, ,\, \,
f_1(q^2) = \frac{0.68 \, GeV^2}{\left(1-\frac{q^2}{30}\right)^2} ~.\nnb
\end{eqnarray}
We present the results of our analysis in a series of
figures. Before the discussion of these figures, we give our SM
predictions for the longitudinal, transverse and the normal
components of the lepton polarizations for \Bgll decay for $\mu$
($\tau$) channel for reference: \bea
<P^-_{L}>  & =  & -0.850 \, (-0.227) \, ,\nnb \\
<P^-_{T}>  & =  & -0.065 \, (-0.190) \, ,\nnb \\
<P^-_{N}>  & =  & -0.014 \, (-0.061) \, .\nnb
\eea
As we noted before, the form factors for B decaying into $\gamma$ and a vacuum state have been also calculated
in the framework of the light front quark model \cite{Geng2} and a model proposed in \cite{Kruger},
which  is based on the constraints   obtained from the  gauge invariance combined with the large energy
effective theory. As reported in  \cite{Kruger} these different approaches for calculating
the form factors causes some uncertain predictions for the branching ratios, in particular, for forward-backward
asymmetries. On the other hand, it seems that the situation with lepton polarization asymmetries in \Bgll decays
is more optimistic since the values of $<P^-_{L}>$, $<P^-_{T}>$ and  $<P^-_{N}>$ given above calculated within
f. eg., the model in \cite{Kruger} turn out to differ only by a small amount, which is less than $10 \%$.

\begin{figure}
\centering
\includegraphics[width=5in]{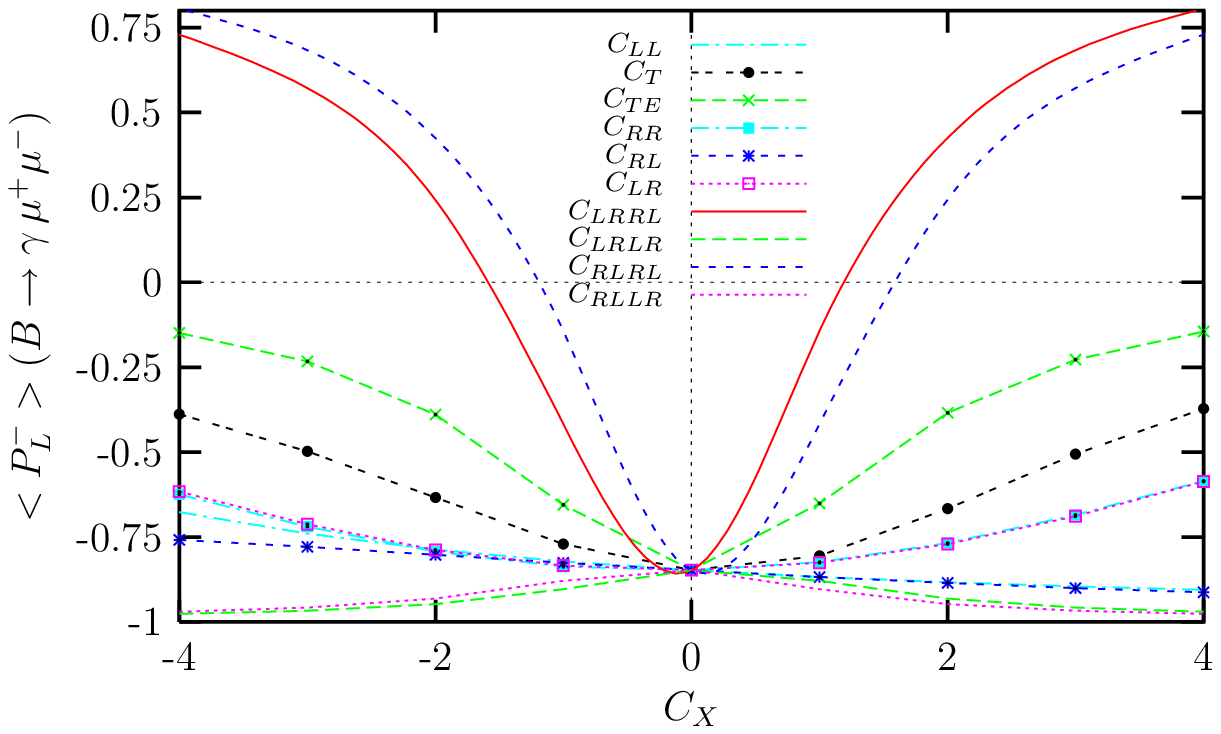}
\caption{The dependence of the averaged longitudinal  polarization $<P^-_L>$ of $\ell^-$ for the
$B_s \rar \gamma \, \mu^+ \mu^-$  decay on the new Wilson coefficients \label{f1}.}
\end{figure}

\begin{figure}
\centering
\includegraphics[width=5in]{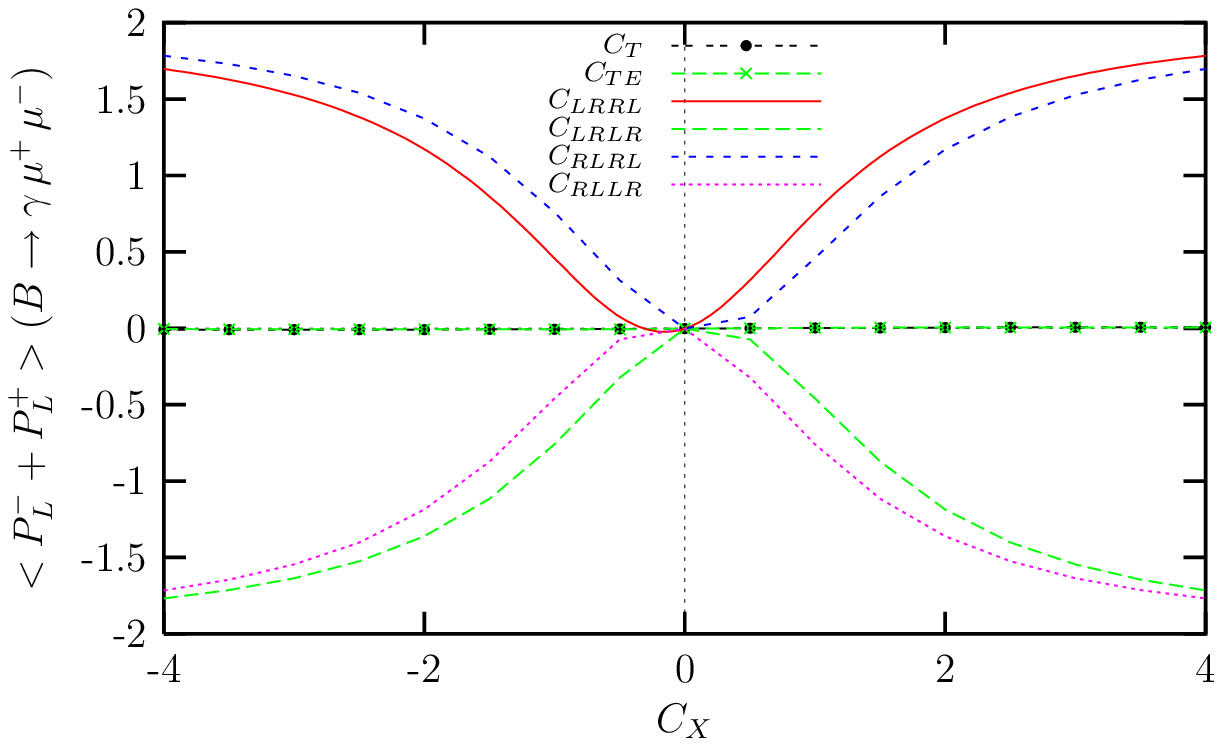}
\caption{The dependence of the combined averaged longitudinal lepton polarization $<P^-_L+P^+_L>$
for the $B_s \rar \gamma \, \mu^+ \mu^-$  decay on the new Wilson coefficients \label{f2}.}
\end{figure}

In Figs. (\ref{f1}) and (\ref{f2}), we present the dependence of the
averaged longitudinal polarization $<P^-_L>$ of $\ell^-$ and the combination $<P^-_L+P^+_L>$
for $B_s \rar \gamma \mu^+ \mu^-$ decay on the new Wilson coefficients. From these figures we see that
$<P^-_L>$ is  strongly dependent on  scalar  type  interactions with coefficient $C_{RLRL}$ and $C_{LRRL}$,
and quite sensitive to the tensor type interactions, while the combined average $<P^-_L+P^+_L>$ is
mainly determined  by scalar  interactions only.  The fact that values of
$<P^-_L>$  becomes substantially different from the SM value (at $C_{X}=0$) as $C_X$ becomes different from zero
indicates that measurement of the longitudinal lepton polarization in $B_s \rar \gamma \mu^+ \mu^-$ decay
can be very useful to investigate new physics beyond the SM.
We note that in Fig. (\ref{f2}), we have not  explicitly exhibit  the
dependence on vector type interactions since we have found that $<P^-_L+P^+_L>$ is not
sensitive them at all. This is what is already expected since vector type interactions are cancelled
when the longitudinal polarization asymmetry of the lepton and antilepton is considered together.
We also observe from Fig. (\ref{f2}) that $<P^-_L+P^+_L>$ becomes almost zero at $C_X=0$, which confirms
the SM result, and its dependence on  $C_X$ is symmetric with respect to this zero point.
It is interesting to note also that $<P^-_L+P^+_L>$ is positive for  all values of  $C_{RLRL}$ and
$C_{LRRL}$, while it is negative for remaining scalar type interactions .

\begin{figure}
\centering
\includegraphics[width=5in]{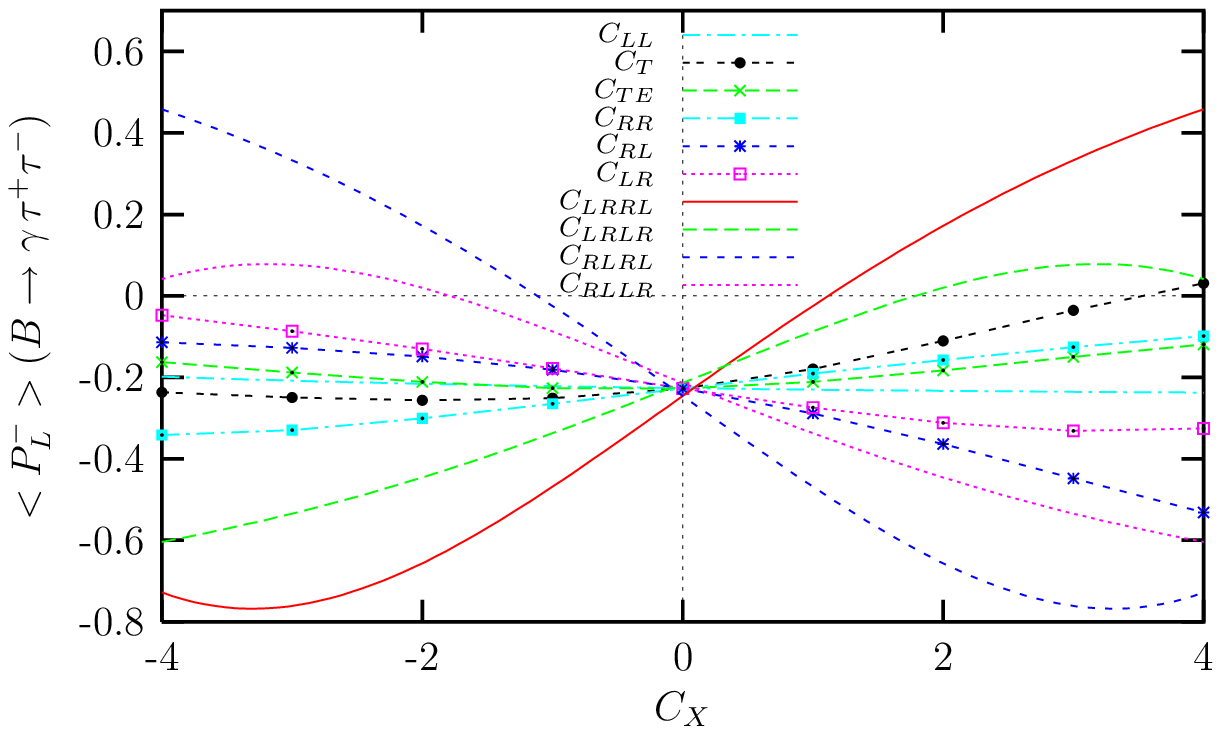}
\caption{The same as Fig.(\ref{f1}), but for the $B_s \rar \gamma \, \tau^+ \tau^-$  decay \label{f3}.}
\end{figure}

\begin{figure}
\centering
\includegraphics[width=5in]{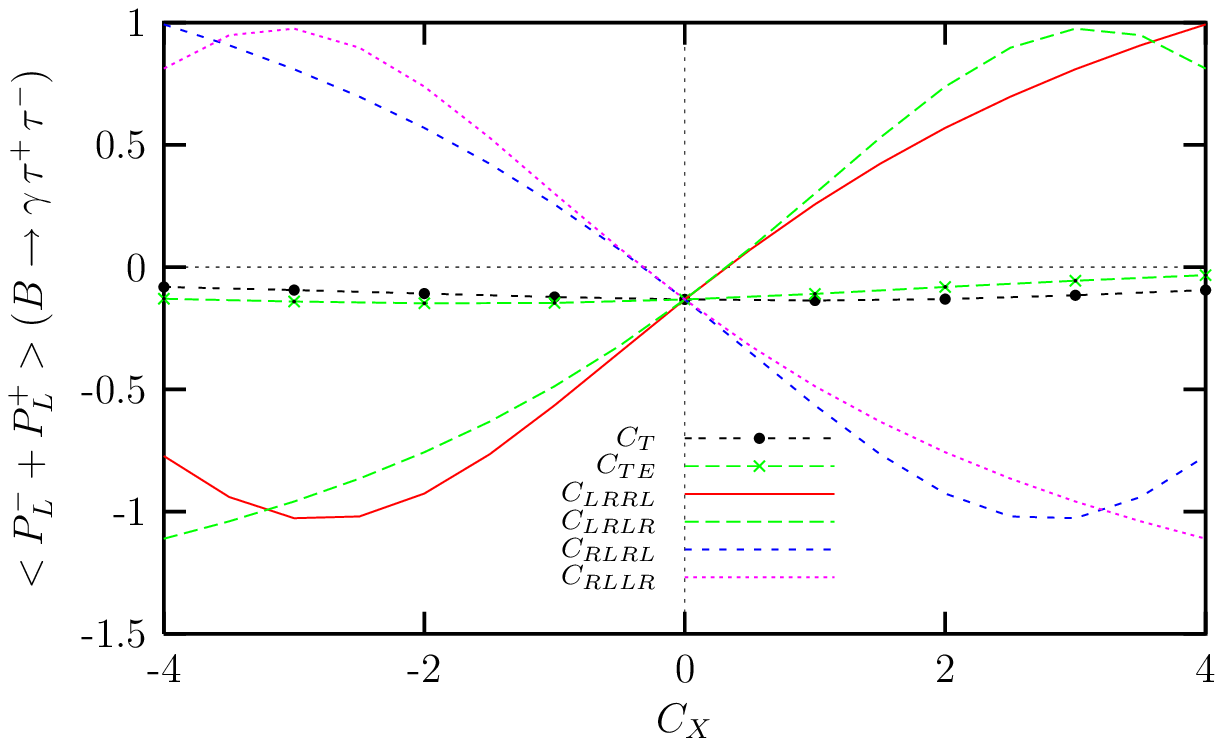}
\caption{The same as Fig.(\ref{f2}), but for the $B_s \rar \gamma \, \tau^+ \tau^-$  decay.\label{f4}}
\end{figure}

Figs. (\ref{f3}) and (\ref{f4}) are the same as Figs. (\ref{f1}) and (\ref{f2}), but for the
$B_s \rar \gamma \tau^+ \tau^-$ decay. Similar to the muon case, $<P^-_L>$ is  sensitive to
scalar type  interactions, but all type. It is an decreasing (increasing) function of
$C_{RLRL}$ and $C_{RLLR}$ ($C_{LRRL}$ and $C_{LRLR}$). The value of $<P^-_L>$ is positive when
$C_{RLRL}\simlt -1$,  $C_{RLLR}\simlt -2$, $C_{LRRL} \simgt 1$ and  $C_{LRLR}\simgt 2$.
As  seen from Fig. (\ref{f4}) that the behavior of
the combined average $<P^-_L+P^+_L>$ for $B_s \rar \gamma \tau^+ \tau^-$ decay is different from the
muon case in that it changes sing for a given scalar type interaction: e.g., $<P^-_L+P^+_L> >0$
when $ C_{RLRL},C_{RLLR} \simlt  0$, while $<P^-_L+P^+_L> <0$
when $C_{RLRL},C_{RLLR}\simgt 0$. Therefore, it can provide valuable information about the new physics
to determine the sign and the magnitude of  $<P^-_L>$ and $<P^-_L+P^+_L>$.

\begin{figure}
\centering
\includegraphics[width=5in]{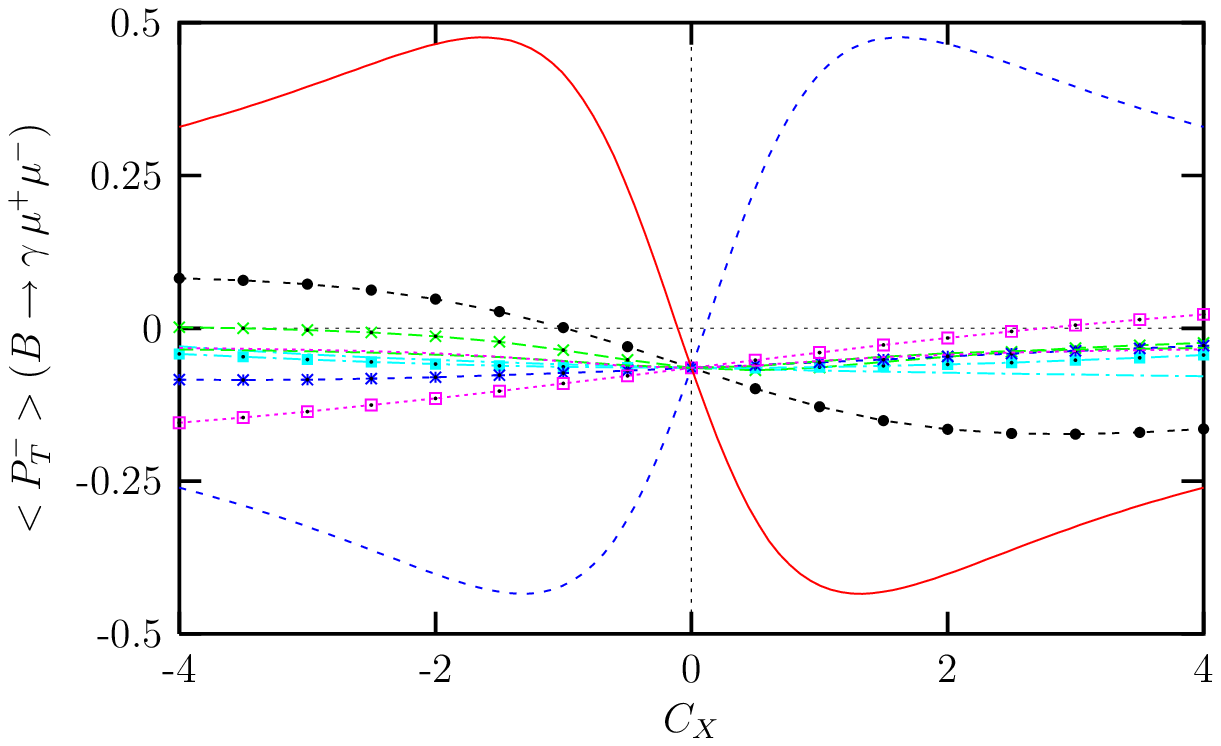}
\caption{The dependence of the averaged transverse  polarization $<P^-_T>$ of $\ell^-$ for the
$B_s \rar \gamma \, \mu^+ \mu^-$  decay on the new Wilson coefficients. The line convention is the
same as before. \label{f5}}
\end{figure}

\begin{figure}
\centering
\includegraphics[width=5in]{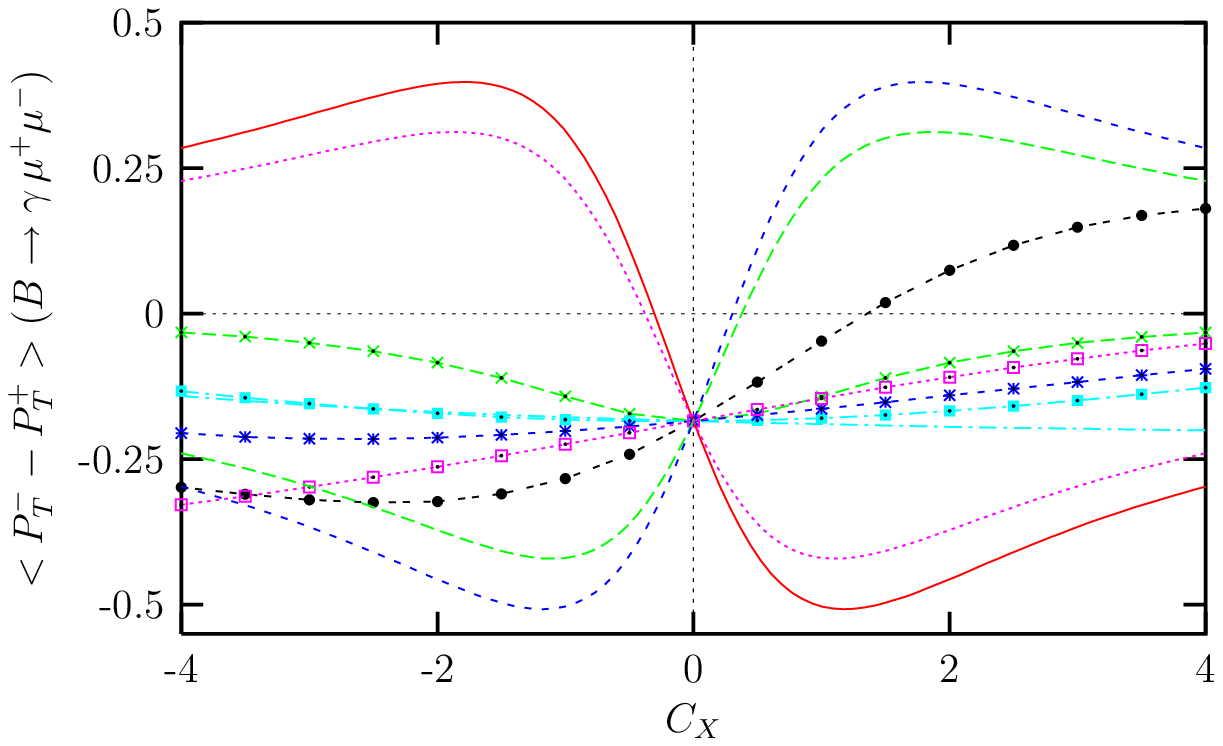}
\caption{The dependence of the combined averaged transverse lepton polarization $<P^-_T-P^+_T>$
for the $B_s \rar \gamma \, \mu^+ \mu^-$  decay on the new Wilson coefficients. The line convention is the
same as before. \label{f6}}
\end{figure}

In Figs. (\ref{f5}) and (\ref{f6}), the dependence of the averaged transverse polarization
$<P^-_T>$ of $\ell^-$ and the combination $<P^-_T-P^+_T>$ for $B_s \rar \gamma \mu^+ \mu^-$ decay
on the new Wilson coefficients are presented. We see from Fig. (\ref{f5}) that $<P^-_T>$
strongly depends on the scalar interactions with coefficient
$C_{RLRL}$ and $C_{LRRL}$ and quite weakly on the all other Wilson coefficients. It is also interesting
to note that $<P^-_T>$ is positive (negative) for the negative (positive) values of $C_{LRRL}$, except
a small region about the zero values of the coefficient, while its behavior with respect to  $C_{RLRL}$
is opposite. As being different from $<P^-_T>$ case, in the combination $<P^-_T-P^+_T>$ there appears
strong dependence on scalar interaction with coefficients $C_{RLLR}$ and $C_{LRLR}$ too, as well as on
$C_{RLRL}$ and $C_{LRRL}$. It is also quite sensitive to the tensor interaction with coefficient $C_T$.

\begin{figure}
\centering
\includegraphics[width=5in]{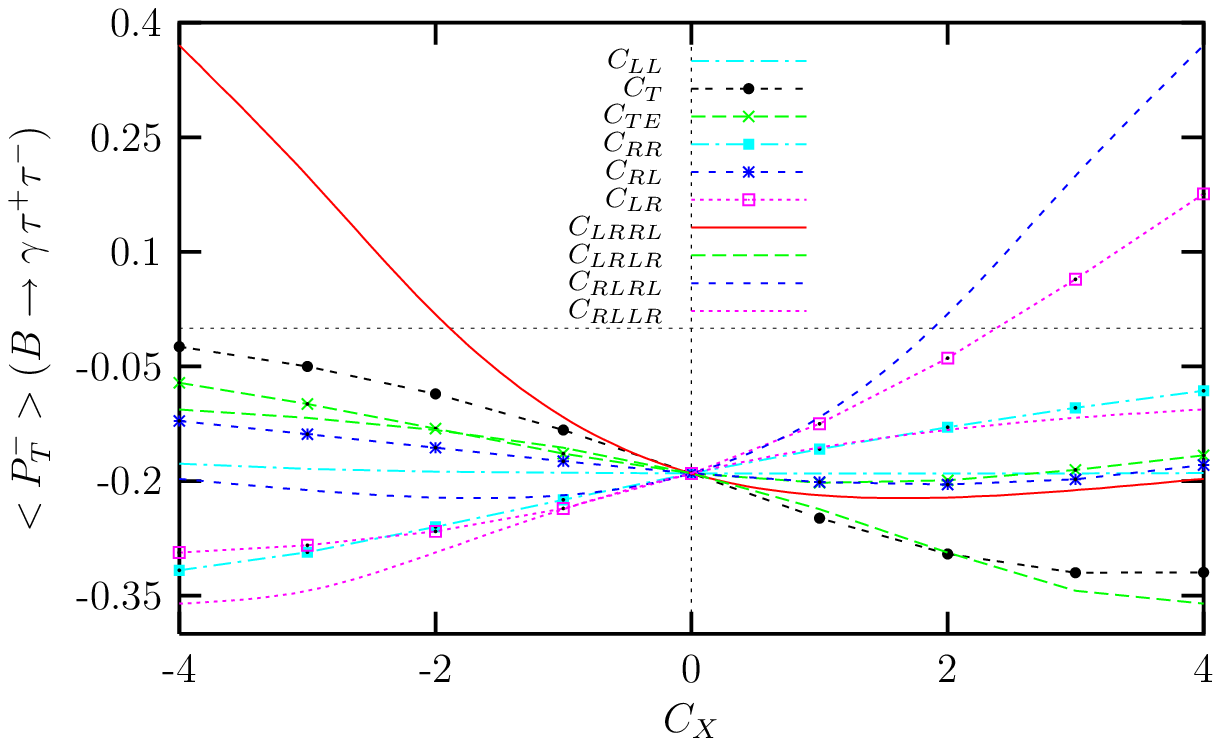}
\caption{The same as Fig.(\ref{f5}), but for the $B_s \rar \gamma \, \tau^+ \tau^-$  decay. \label{f7}}
\end{figure}

\begin{figure}
\centering
\includegraphics[width=5in]{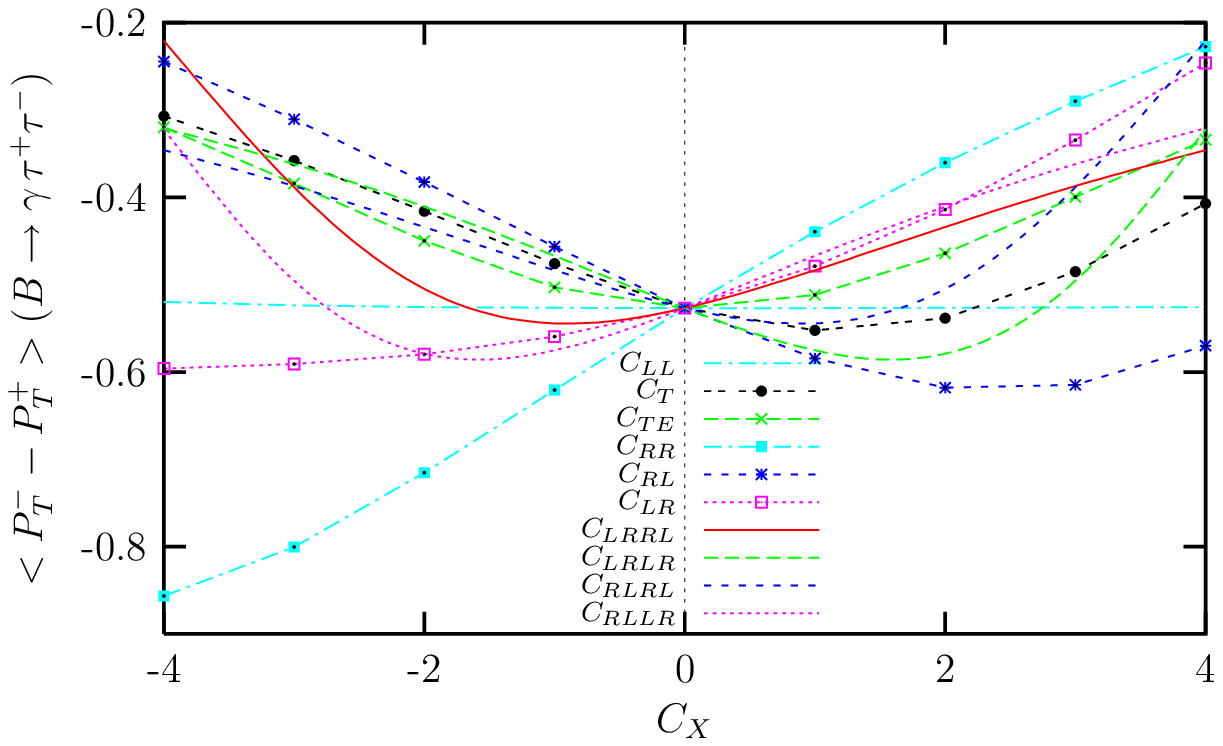}
\caption{The same as Fig.(\ref{f6}), but for the $B_s \rar \gamma \, \tau^+ \tau^-$  decay.\label{f8}}
\end{figure}

Figs. (\ref{f7}) and (\ref{f8}) are the same as Figs. (\ref{f5}) and (\ref{f6}), but for the
$B_s \rar \gamma \tau^+ \tau^-$ decay. As in the muon case, for $\tau$ channel too, the dominant contribution
to the transverse polarization comes from the scalar interactions, but it exhibits a more sensitive
dependence to the remaining types of interactions as well than the muon case. As seen from Fig. (\ref{f8})
that $<P^-_T-P^+_T>$ is negative for all values of the new Wilson coefficients, while $<P^-_T>$ again
changes sign depending on the change in the new Wilson coefficients: e.g., $<P^-_T> >0$ only when
$C_{LRRL}\simlt -2$ and $C_{RLRL},C_{LR}\simgt 2$. Remembering that in SM in massless lepton case, $<P^-_T>\approx 0$
and  $<P^-_T-P^+_T> \approx 0$, determination of the sign of these observables can give useful information
about the existence of new physics.

\begin{figure}
\centering
\includegraphics[width=5in]{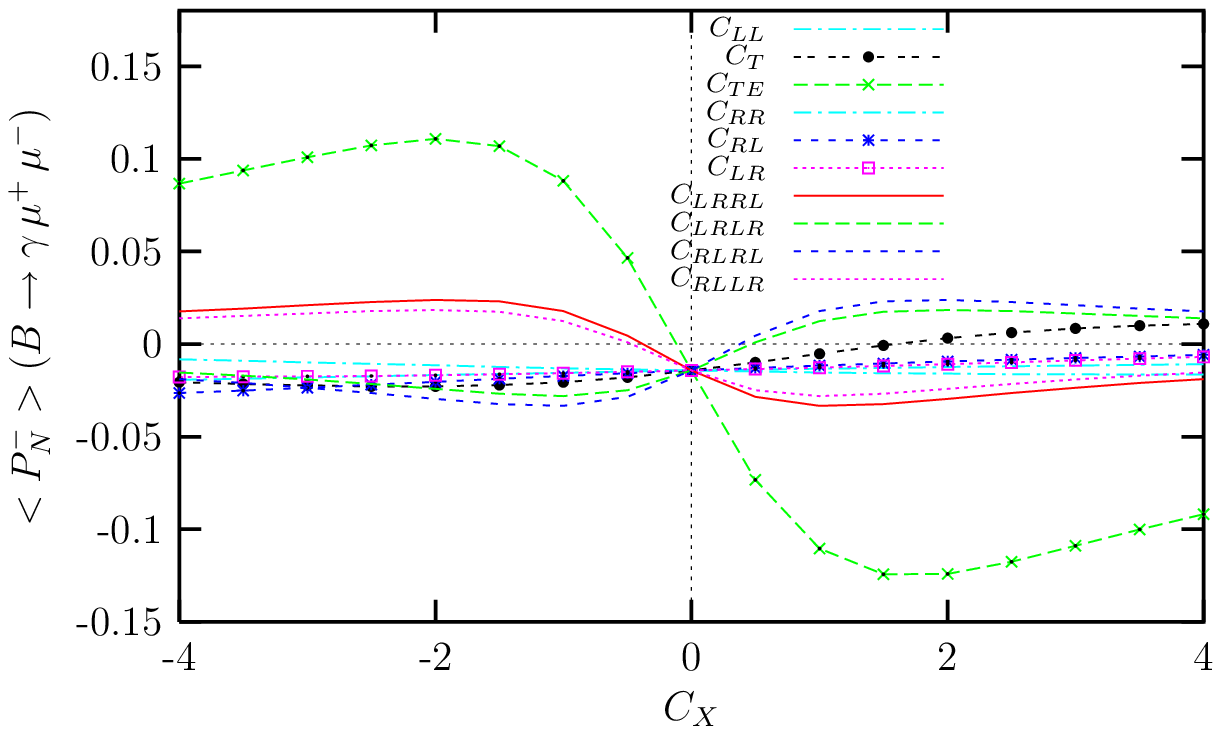}
\caption{The dependence of the averaged normal polarization $<P^-_N>$ of $\ell^-$ for the
$B_s \rar \gamma \, \mu^+ \mu^-$  decay on the new Wilson coefficients \label{f9}.}
\end{figure}

\begin{figure}
\centering
\includegraphics[width=5in]{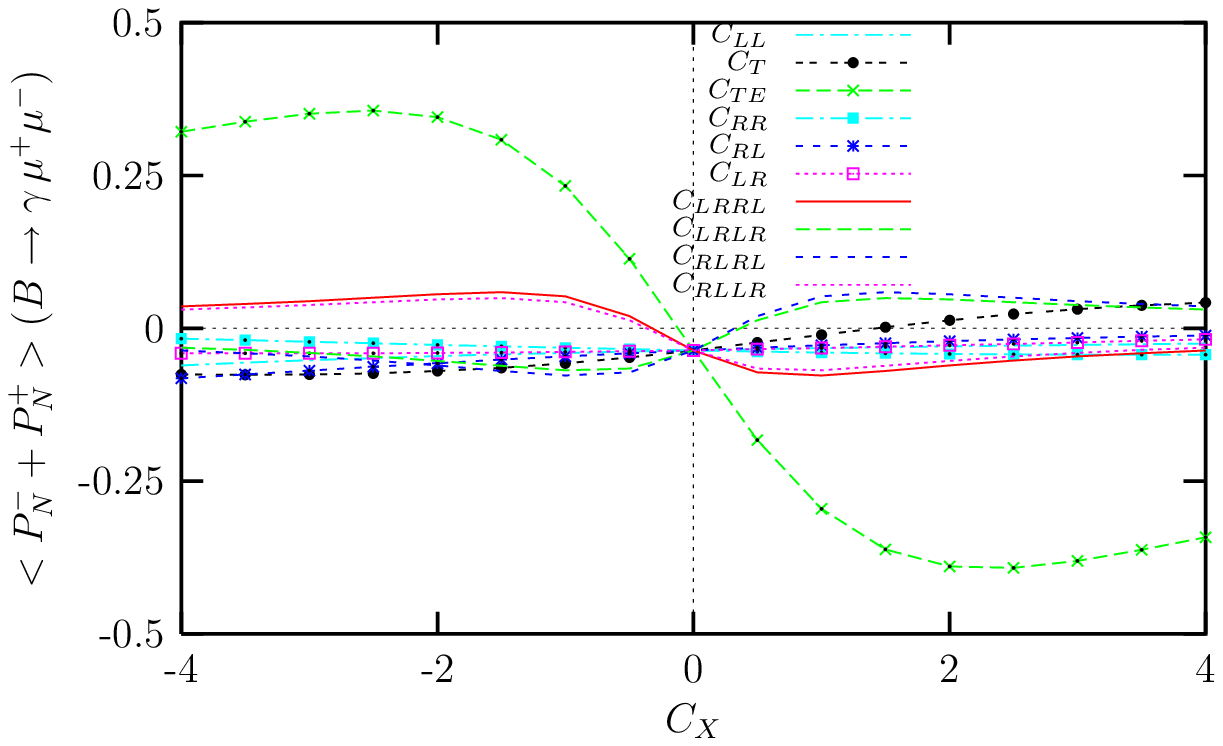}
\caption{The dependence of the combined averaged normal lepton polarization $<P^-_N+P^+_N>$
for the $B_s \rar \gamma \, \mu^+ \mu^-$  decay on the new Wilson coefficients.\label{f10}}
\end{figure}

In Figs. (\ref{f9}) and (\ref{f10}), we present the dependence of the
averaged normal polarization $<P^-_N>$ of $\ell^-$ and the combination $<P^-_N+P^+_N>$
for $B_s \rar \gamma \mu^+ \mu^-$ decay on the new Wilson coefficients. We observe from these figures
that behavior of  both $<P^-_N>$ and $<P^-_N+P^+_N>$ are determined  by  tensor type interactions
with coefficient $C_{TE}$. They both are  positive (negative) when  $C_{TE}\simlt 0 $ ($C_{TE}\simgt 0 $).

\begin{figure}
\centering
\includegraphics[width=5in]{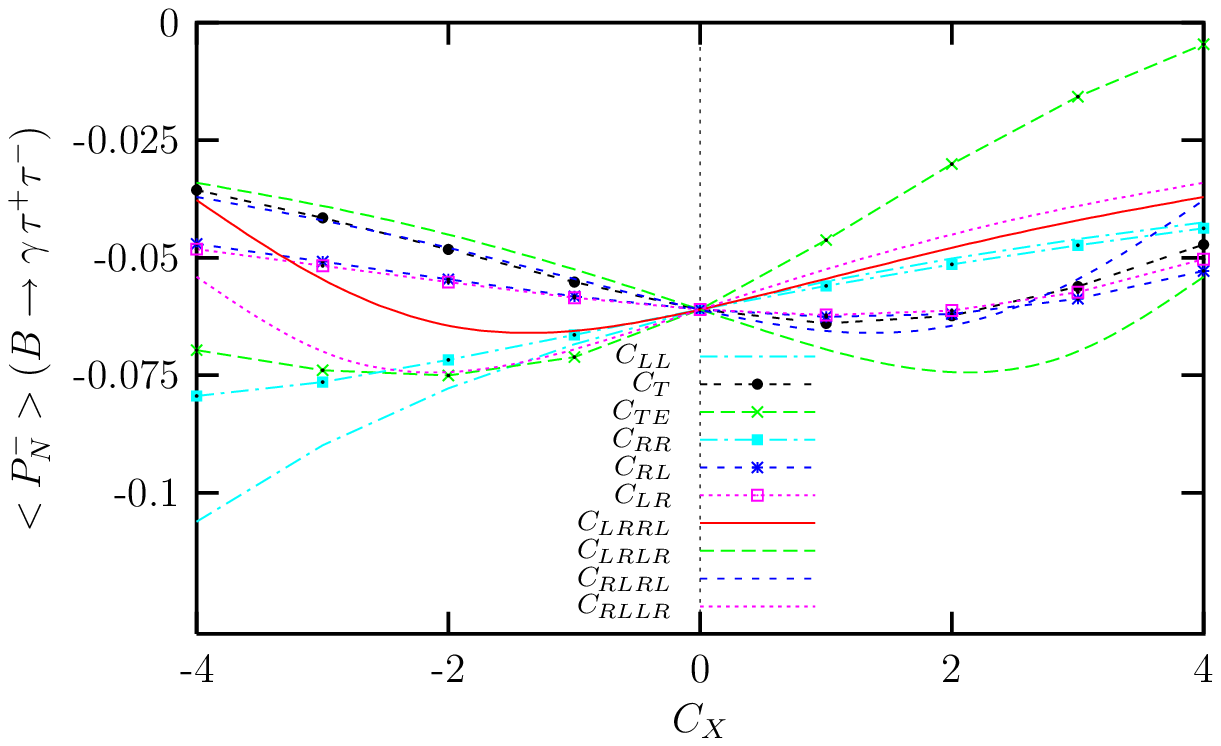}
\caption{The same as Fig.(\ref{f9}), but for the $B_s \rar \gamma \, \tau^+ \tau^-$  decay.\label{f11}}
\end{figure}

\begin{figure}
\centering
\includegraphics[width=5in]{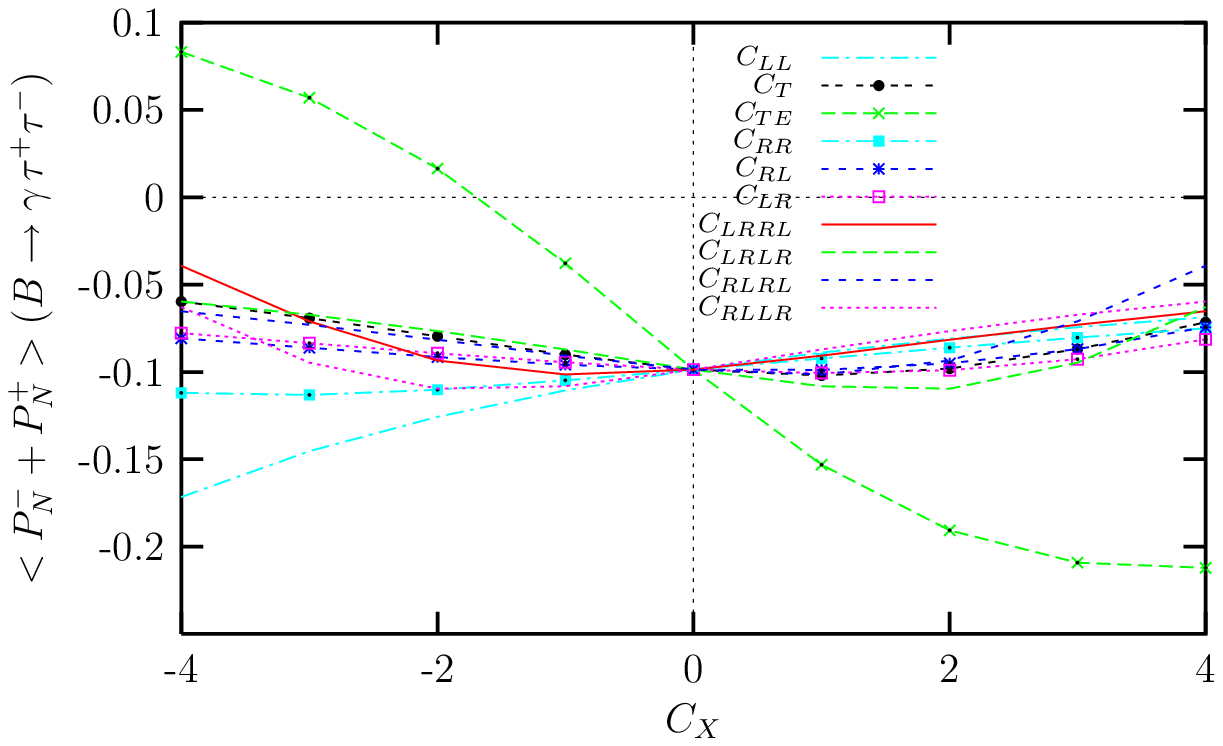}
\caption{The same as Fig.(\ref{f10}), but for the $B_s \rar \gamma \, \tau^+ \tau^-$  decay.\label{f12}}
\end{figure}

Figs. (\ref{f11}) and (\ref{f12}) are the same as Figs. (\ref{f9}) and (\ref{f10}), but for the
$B_s \rar \gamma \tau^+ \tau^-$ decay. As being different from the muon case,  $<P^-_N>$ for $\tau$ channel
is also sensitive to the vector type interaction with coefficient $C_{LL}$, as well as the tensor types
and it is negative for all values of the new Wilson coefficients.  As for  the combination $<P^-_N+P^+_N>$
for $\tau$ channel, it is negative too for  all values of $C_X$, except for $C_{TE}\simlt -2$.

We now summarize our results:
\begin{itemize}
\item $<P^-_L>$ and $<P^-_T>$ are  strongly dependent on  scalar  type  interactions with coefficient $C_{RLRL}$
and $C_{LRRL}$, while $<P^-_N>$ is mainly determined  by  tensor type interactions
with coefficient $C_{TE}$.
\item Measurement of $<P^-_L>$ in $B_s \rar \gamma \mu^+ \mu^-$ decay
can be very useful to investigate new physics beyond the SM since it
 becomes substantially different from the SM value (at $C_{X}=0$) as $C_X$ becomes different from zero.
\item The combined averages $<P^-_L+P^+_L>$ and $<P^-_T-P^+_t>$  are mainly determined  by scalar  interactions
only. As for the  $<P^-_N+P^+_N>$, it is quite sensitive to tensor type interactions with coefficient $C_{TE}$.
\item $<P^-_L+P^+_L>$ becomes almost zero at $C_X=0$, which confirms
the SM result and it is  positive for  all values of  $C_{RLRL}$ and
$C_{LRRL}$, while it is negative for remaining scalar type interactions.
\item Since in the SM in massless lepton case, $<P^-_T>\approx 0$
and  $<P^-_T-P^+_T> \approx 0$, determination of the sign of these observables can give useful information
about the existence of new physics.
\end{itemize}

In conclusion, we have   studied the lepton polarizations in
the rare $B_s \rar \gamma \,\ell^+ \ell^- $ decays  by using the general, model independent form
of the effective Hamiltonian. The sensitivity of the longitudinal, transverse and normal polarizations
of $\ell^-$, as well as lepton-antilepton combined asymmetries, on the new Wilson coefficients are
investigated. We find that all these physical observables  are very sensitive to the
existence of new physics beyond SM and their experimental measurements can give valuable information
about it.

\newpage
\renewcommand{\topfraction}{.99}
\renewcommand{\bottomfraction}{.99}
\renewcommand{\textfraction}{.01}
\renewcommand{\floatpagefraction}{.99}

\clearpage
\clearpage
\clearpage
\clearpage
\clearpage
\end{document}